# Thermal Radiation Effect in the Free Expansion of an Ideal Gas and Gibbs' Paradox in Classical Thermodynamics

A. Paglietti

Department of Structural Engineering
University of Cagliari, 09123 Cagliari, Italy
*e-mail:* paglietti@unica.it

**Abstract:** The standard theory of ideal gases ignores the interaction of the gas particles with the thermal radiation (*photon gas*) that fills the otherwise vacuum space between them. This is an unphysical feature of the theory since every material in this universe, and hence also the particles of a gas, absorbs and radiates thermal energy. The interaction with the thermal radiation that is contained within the volume of the body may be important in gases since the latter, unlike solids and liquids, are capable of undergoing conspicuous volume changes. Taking this interaction into account makes the behaviour of the ideal gases more realistic and removes Gibbs' paradox.

## 1. Introduction

A rigid vessel is composed of two equal chambers separated by a removable partition. The two chambers are filled with two different ideal gases at the same initial temperature $T_0$. Let $n_1$ and $n_2$ be the number of moles of gas contained in the first and in the second chamber, respectively. The gases in the vessel cannot exchange heat with the surroundings as the vessel walls are adiabatic. As the partition is removed, which can be done without performing any work on the gases, the two gases diffuse through each other. This is a spontaneous irreversible process and we are interested in determining the entropy increase it produces.

The problem is a standard one in Classical Thermodynamics. Its solution can be found in many textbooks. In particular, the excellent treatise by Fast [1, pp. 41-43] contains a clear approach to this problem, entirely within the realm of Classical

Thermodynamics. Being ideal, the two gases do not interact with each other during the mixing process. This means, in particular, that as the partition is removed each of them expands into the volume occupied by the other, as if it expanded in a vacuum. Furthermore, since the free expansion of an ideal gas in a vacuum does not produce any temperature change, the final temperature of the gas mixture must be the same as the initial temperature $T_0$. The entropy change $\Delta S$ due to the considered expansion process is obtained directly from the classical expression of the entropy of an ideal gas:

$$S = n\,(c_v \ln T + R \ln v) + \text{const.} \qquad (1.1)$$

Here $c_v$ is the molar specific heat at constant volume, $n$ the number of moles of the gas, $R$ the universal gas constant, $T$ the absolute temperature, while $v = V/n$ is the volume $V$ per mole. It is thus found that

$$\Delta S = R\,(n_1 \ln \frac{V_1 + V_2}{V_1} + n_2 \ln \frac{V_1 + V_2}{V_2}). \qquad (1.2)$$

where $V_1$ and $V_2$ are the volumes of the two chambers. In the present case we have $V_1 = V_2 = V$, which simplifies eq. (1.2) to:

$$\Delta S = R\,(n_1 + n_2)\ln 2. \qquad (1.3)$$

The so-called *Gibbs' paradox* arises because the above formulae are independent of the physicochemical nature of the two gases. This seems hardly acceptable. Surely, the entropy change resulting from mixing together say one mole of helium and one mole of ammonia should be different than that resulting from mixing two similar amounts of two different isotopes of oxygen. In other words a change in the properties of the mixing gases should produce a change in the entropy increase due to the mixing, which is denied by the above formulae.

The same formulae apply in particular when $n_1 = n_2 = 1$. In this case eq. (1.3) simplifies to

$$\Delta S = 2R \ln 2 > 0. \qquad (1.4)$$

Now, if the two ideal gases are the same, the gas pressure will be the same in the two chambers because they both contain one mole of the gas at the same volume and the same temperature. In this case too the theory predicts the same entropy change (1.4) following the partition removal. No entropy increase should however occur, simply because no change in the state of the two gases takes place as the partition is removed. This inconsistency in itself is often referred to as the Gibbs' paradox. Actually, it is a further consequence of the fact that eqs (1.2) and (1.3) do not depend on the properties of the mixing gases.

The existence of Gibbs' paradox casts some shadows on the classical theory of ideal gases. It suggests that that theory may somehow be flawed even in the range of pressures and temperatures where it turns out to provide an otherwise superb approximation of the behaviour of the real gases.

The same paradox is met when the theory of ideal gases is approached by the methods of Statistical Mechanics. In that case the paradox is traditionally resolved in various ways. The most convincing one introduces the notion of indistinguishability of the various molecules of the same gas when calculating the number of the available micro-states. Another, though more summary way to fix the paradox is to dismiss as meaningless any calculation about the mixing of two identical gases, on the ground that in this case there are not actually two different gases to mix. Both ways only remove the paradox in the case in which the mixing gases are the same, though. When it comes to two different gases, also the Statistical Mechanics approach predicts that the entropy change due to their mixing should not depend on the nature of the gases. A way out of this shortcoming has been proposed by Lin. For a recent commentary of his proposal and reference to the original papers the reader is referred to [2]. Lin's approach, however, introduces the concept of information entropy, which is foreign to Classical Thermodynamics.

In looking for a solution to this problem, one cannot help but observe that Classical Thermodynamics and Statistical Mechanics are essentially two different approaches, resting on entirely different bases. Each approach should therefore be consistent within its own framework. It is only a matter of scientific rigour then, that we should not use one approach to justify the other. For this reason, in the following sections we shall attempt to resolve the various aspects of Gibbs' paradox within the realm of Classical Thermodynamics.

More precisely, we shall show that the origin of the paradox lies in the fact that the notion of ideal gas, as introduced in Classical Thermodynamics or, for that matter, also in Statistical Mechanics, cannot be entirely realistic. This is so because it represents a material that does not radiate any energy. Every material in this universe radiates energy, depending on its temperature. This fact is ignored by the theory of the ideal gases. The latter are imagined as being made of volumeless particles, endowed with mass, unable to interact at distance with each other and with the radiation that always fills the very space in which they are moving about. As a matter of fact, that space will exchange energy with them through thermal radiation until thermal equilibrium is reached. Taking into account this phenomenon will make the ideal gas more physical and also remove Gibbs' paradox.

A similar interaction can be neglected in liquids and solids, since they undergo only minor volume changes.

## 2. The photon gas that fills the vacuum amid the gas particles

Any cavity or otherwise empty space harbours electromagnetic radiation within it. This radiation is often referred to as *thermal radiation* or *photon gas*. The latter terminology is particularly appropriate because from the macroscopic standpoint such a radiation behaves in many ways as a gas. In this section the main formulae concerning the photon gas are recalled, as can be found in many books of Classical Thermodynamics (see e.g. [1, pp.158-162] or [3, Sect. 13.16]). They apply in thermal equilibrium conditions, that is when the walls of the cavity and/or any material particle within it have reached the same equilibrium temperature. The latter will be referred to as the temperature of the photon gas itself. Its value in the absolute temperature scale will be denoted by $T$, as usual.

In thermal equilibrium conditions, a unit volume of space filled with a photon gas stores the energy $u$ given by

$$u = aT^4. \qquad (2.1)$$

The constant $a$ appearing here can be expressed as

$$a = \frac{4}{c}\sigma, \qquad (2.2)$$

where $c$ is the velocity of light while $\sigma$ is the well-known Stefan-Boltzman constant [$\sigma = 56.697$ nW/m$^2$ deg$^4$, which makes $a=75.646\ 10^{-8}$ nJ/m$^3$deg$^4$ since $c=2.998\ 10^8$ m/sec]. The internal energy $U$ of a volume $V$ of photon gas is accordingly given by

$$U = aT^4 V. \qquad (2.3)$$

Under the same conditions, the photon gas exerts a pressure $P$ on the walls of the cavity, the value of which is

$$P = \frac{1}{3}u = \frac{1}{3}aT^4. \qquad (2.4)$$

This pressure does work as the volume expands or contracts. The work supplied by the photon gas to the cavity walls as its volume is increased by d$V$ turns out to be

$$\mathrm{d}W_{out} = P\,\mathrm{d}V = \frac{1}{3}u\,\mathrm{d}V = \frac{1}{3}aT^4\,\mathrm{d}V. \qquad (2.5)$$

Thus if $Q$ denotes the amount of heat absorbed by the photon gas, we can apply the 1$^{st}$ principle of Thermodynamics to state that

$$dU = dQ - dW_{out}. \quad (2.6)$$

From this and from eqs (2.3) and (2.4) we get

$$dQ = 4\,aT^3 V\,dT + \frac{4}{3}aT^4\,dV. \quad (2.7)$$

By applying this equation to a reversible process we obtain

$$dS = 4\,aT^2 V\,dT + \frac{4}{3}aT^3\,dV, \quad (2.8)$$

since $dQ = TdS$ for reversible processes. The total differential equation (2.8) can easily be integrated to give the following expression for the entropy $S$ of the photon gas:

$$S = \frac{4}{3}aT^3 V, \quad (2.9)$$

the integration constant being set equal to zero, since $S = 0$ for $T = 0$.

Finally, from eq. (2.7) specific (per unit volume) thermal capacity $r_v$ of the photon gas at constant volume is immediately obtained:

$$r_v = \frac{1}{V}\frac{dQ}{dT} = 4\,aT^3. \quad (2.10)$$

## 3. Influence of thermal radiation on the adiabatic free expansion of an ideal gas

According to classical theory, a free adiabatic expansion of an ideal gas should leave its temperature unaltered. This follows from the fact that the particles of the gas are free from long-range interactions, which means that they can only store kinetic energy. In a free expansion no external work is done by the gas, since it expands in a vacuum. No thermal energy is exchanged either, as the expansion is adiabatic. The kinetic energy of the gas is therefore conserved, which in particular means that the gas temperature in the final equilibrium state after the expansion must be the same as the initial one.

The above explanation does not take into account that no material is physically admissible if it does not emit and absorb thermal radiation. The ideal gas is lacking in this respect, which may lead to some physical inconsistencies in its behaviour. In the present section we shall show how the interaction with the ubiquitous photon gas, which cannot be avoided by any real gas –not even in the ranges of temperatures and pressures where its state equations coincide with those of an ideal gas– makes the gas cool as a result of a free adiabatic expansion.

Let us first of all determine the amount of heat that a photon gas absorbs from the surroundings as it expands at constant temperature. If $\Delta V$ is the volume increase of the gas, the amount of heat needed to expand it isothermally is given by

$$\Delta Q = \frac{4}{3} a T^4 \Delta V . \quad (3.1)$$

This immediately follows from eq. (2.7) once we set d$T$=0 and integrate the resulting equation between the initial volume $V$ and the final one $V + \Delta V$.

If this amount of heat is not supplied, which is the case when the expansion takes place adiabatically, then the volume change $\Delta V$ will produce a reduction in the photon gas temperature. On the other hand, if the cavity walls are adiabatic but the cavity itself also contains a gas, then the amount of heat $\Delta Q$ will be taken from the particles of the gas, which will cool down as a result. In this case, the temperature change can to a good approximation be calculated by assuming that the thermal capacity of the photon gas is negligible with respect to that of the ideal gas within the cavity. Keeping in mind that the molar specific heat $c_v$ of an ideal gas does not depend on volume and referring to the case in which the cavity contains just one mole of ideal gas, the temperature variation due to the expansion of the latter will be given by

$$\Delta T = -\frac{\Delta Q}{c_v} = -\frac{4}{3} \frac{a T^4 \Delta V}{c_v} . \quad (3.2)$$

If there are $n$ moles of gas in the cavity, this temperature variation should be divided by $n$. In writing eq. (3.2) we assumed that $\Delta T$ was small enough as to produce a negligible change in $c_v$ during the expansion. This is certainly so in the vast majority of cases. In any case, taking account of the dependence of $c_v$ on $T$ does not appear to pose any serious problem.

The entropy change of the system due to the considered expansion is the sum of the contribution $\Delta S_g$ coming from the ideal gas and the contribution $\Delta S_r$ of the photon gas. The former is readily obtained from eq. (1.2) and is given by

$$\Delta S_g = c_v \ln \frac{T + \Delta T}{T} + R \ln \frac{V + \Delta V}{V} \quad (3.3)$$

for each mole of gas. The other contribution is calculated from eq. (2.8) to be given by:

$$\Delta S_\mathrm{r} = 4\,aT^2 V\,\Delta T + \frac{4}{3}aT^3\,\Delta V\,. \tag{3.4}$$

If the volume *V* is not too large, $\Delta S_\mathrm{r}$ can be neglected with respect to $\Delta S_\mathrm{g}$. Moreover, for small values of $\Delta T$ as the one involved in the present case, the first term in the right-hand side of eq. (3.4) can be neglected too, as long as *T* is sufficiently far from zero. Under these conditions we can, to a good approximation, calculate the total entropy change of the system as

$$\Delta S = R\ln\frac{V+\Delta V}{V}\,. \tag{3.5}$$

In conclusion, we can say that as far as the considered process is concerned the entropy change of the system is not essentially affected by the presence of the photon gas. The latter, however, does produce a cooling effect in the ideal gas temperature resulting after the expansion. The measure of this effect is given by eq. (3.2).

**4. The entropy change of the two gases ensuing from the partition removal**

Let us now refer to the two gases filling the two separate chambers of the adiabatic vessel considered in Sect. 1. As the partition is removed, the two gasses expand by flowing one through the other. Each gas expands as if it were in a vacuum, since the two gases are supposed to be ideal. This enables us to calculate the entropy change of the system of the two gases by referring to the following sequence of processes:

(1) A free adiabatic expansion of each separate gas, bringing them to the same final volume 2*V*. In this process, the gas does not supply/absorb any work to/from the surroundings since the process can be assimilated to a free expansion in a vacuum. The same process will produce a change in the entropy and temperature of each expanding gas, though. Moreover, different gases will suffer different temperature changes.

(2) A transfer of heat between the two expanded gases bringing them to the same final temperature T*. Being a heat transfer from a hotter gas to a colder one, this process will result in a further entropy change of the system. In the process, no heat is exchanged with the surroundings as the vessel walls are adiabatic.

(3) Finally, the two gases are allowed to mix reversibly at the constant volume 2*V* they reached at the end of phase (1). A process of this kind is admissible from the physical standpoint and was first conceived by Plank [4, Sect. 236] (see also [1, p. 41] and [5, p. 729]). Being adiabatic and reversible, the process will not produce any entropy change. It will not produce any change in the temperature of the gases either, since the latter are ideal and the process does not bring about any volume change. The system will thus reach the final state in which the two gases are mixed together at temperature T* and volume 2V.

The details of each of the above three phases of the process are worked out below.

**Phase 1.** (*Free adiabatic expansion*)

One mole of an ideal gas fills the volume $V$ of one chamber of the vessel. Let $T_0$ be its initial temperature and $c'_v$ its molar specific heat at constant volume. The gas expands adiabatically in a vacuum until it occupies the volume $2V$ of the entire vessel, thus increasing its volume by $\Delta V = V$. If the presence of heat radiation is taken into account, the equilibrium temperature $T_1$ that results after the expansion can be calculated from eq. (3.2) to be given by:

$$T_1 = T_0 - \frac{4}{3} \frac{a T_0^4 \Delta V}{c'_v} \ . \tag{4.1}$$

A similar adiabatic expansion of the gas filling the other chamber will bring it to the final volume $2V$ and to a final equilibrium temperature $T_2$ given by:

$$T_2 = T_0 - \frac{4}{3} \frac{a T_0^4 \Delta V}{c''_v} \ , \tag{4.2}$$

where $c''_v$ is the molar specific heat at constant volume of this new gas.

In order to calculate the entropy change brought about by this phase of the process, we first observe that if $\Delta V$ is not too large, both $T_1$ and $T_2$ will be sufficiently near to $T_0$ as to allow us to apply the approximate equation (3.5). This process will therefore increase the entropy of each gas by the amount

$$\Delta_e S = R \ln 2. \tag{4.3}$$

This will produce an overall increase in the entropy of the two gases by the amount

$$\Delta_1 S = 2 \Delta_e S = 2 R \ln 2. \tag{4.4}$$

It is apparent that such an entropy change coincides with that which would be expected for an ideal gas in the absence of the photon gas. The presence of the latter, however, will produce a cooling effect that is proportional to $\Delta V$ resulting from the considered expansion, as predicted by eqs (4.1)-(4.2). Small as this effect may be (it is proportional to $\Delta V$, though!), it cannot be avoided not even by an ideal gas, since every material must absorb and radiate thermal energy. Most importantly, the amount of cooling depends on the specific heat of the gas, so that it will generally be different for different ideal gases.

**Phase 2.** (*Internal heat transfer*)

The previous process brings the two gases to states ($T_1$, $2V$) and ($T_2$, $2V$), respectively. To be definite, we shall assume that $c'_v > c''_v$, which in view of eqs (4.1) and (4.2) implies that $T_1 > T_2$. We now put the two gases in thermal contact with each other while keeping them thermally insulated from the surroundings. In these conditions the hotter gas will supply heat to the colder one until both gases reach the same equilibrium temperature $T^*$. The determination of $T^*$ is an elementary problem of heat transfer. It is solved by equating the total amount of heat $Q_1^*$ lost by the hotter gas to the amount of heat $Q_2^*$ absorbed by the colder one. Since we are considering one mole of each gas, we have

$$Q_1^* = (T_1 - T^*)\, c'_v \qquad (4.5)$$

and

$$Q_2^* = (T^* - T_2)\, c''_v. \qquad (4.6)$$

We therefore get

$$T^* = \frac{T_1 c'_v + T_2 c''_v}{c'_v + c''_v}. \qquad (4.7)$$

Once the amounts of heat (4.5) and (4.6) are known, the entropy change caused by the process can also be obtained. Since the temperatures $T_1$, $T_2$ and $T^*$ are close together, we shall not introduce any serious mistake if we assume that the heat $Q_1^*$ is lost at the constant temperature $T_1$ and that the heat $Q_2^*$ is absorbed at constant temperature $T_2$. The entropy change due to the heat transfer is, accordingly:

$$\Delta_2 S = -\frac{Q_1^*}{T_1} + \frac{Q_2^*}{T_2} = \frac{c'_v c''_v}{c'_v + c''_v} \frac{(T_1 - T_2)^2}{T_1 T_2}, \qquad (4.8)$$

which is clearly greater than zero. The important point to be noted here is that this entropy change depends on the specific heats of the mixing gases. Such a dependence is both explicit, as shown by equation (4.8), and implicit through $T_1$ and $T_2$ via equations (4.1) and (4.2). In the case in which $c'_v = c''_v$, the latter equations yield $T_1 = T_2$, which makes $\Delta_2 S = 0$. In particular, $\Delta_2 S$ vanishes if the two chambers are filled with the same gas.

**Phase 3.** (*Isothermal reversible mixing*)

As previously observed, in this phase of the process the two perfect gases do not suffer any change in entropy or temperature since they undergo a reversible isothermal mixing. The presence of the photon gas does not alter this conclusion, since this phase of the process takes place at a constant volume.

From eqs (4.4) and (4.8) the entropy change due to the whole process turns out to be, therefore:

$$\Delta S = \Delta_1 S + \Delta_2 S = 2 R \ln 2 + \frac{c'_v c''_v}{c'_v + c''_v} \frac{(T_1 - T_2)^2}{T_1 T_2}. \qquad (4.9)$$

The first term in the far right-hand side of this equation is due to the volume expansion and has the same value no matter the ideal gases under consideration. The last term attains different values for different mixing gases.

Incidentally, such a result helps to define precisely when two ideal gases are to be treated as the same or as different as far as the thermodynamics of their mixing is concerned. It shows that it all depends on whether they have the same specific heat or not. Other differences in their properties are not relevant from the thermodynamical standpoint. This fact should be taken into due account when a Statistical Mechanics approach to this phenomenon is sought. It implies that some otherwise different microstates of a gas mixture should be considered to be the same if they are relevant to two different ideal gases which possess the same specific heat.

## 5. The solution of Gibb's paradox

According to the analysis of the previous section, there are two contributions to the entropy change $\Delta S$ following the removal of the partition between the two ideal gases contained in the adiabatic vessel we considered in the Introduction. The first contribution is $\Delta_1 S$ and is relevant to what we called Phase 1 of the process. It is due to the separate expansion of each gas and is practically independent of their particular physicochemical properties [cf. eq. (4.3)]. The crucial point to be noted here is, however, that if account of thermal radiation is duly taken, the final temperature resulting from this phase of the process does depend on the properties of the expanding gas; more precisely, on its specific heat. This means that it is different for different gases [cf. eqs. (4.1) and (4.2)].

The other contribution to the entropy change, namely $\Delta_2 S$, results from Phase 2. At a variance with $\Delta_1 S$, this contribution depends on the specific heats $c'_v$ and $c''_v$, of the two gases. Moreover, as the difference between $c'_v$ and $c''_v$ tends to zero, so does $\Delta_2 S$. As already observed, this shows that as far as this phase of the process is concerned, a quantitative measure of the difference between the two gases is provided by the difference in their specific heats. Any property of the gases other than their specific heats has no effect on the entropy change resulting from this part of the process.

From the above results we can conclude that the total entropy change $\Delta S$ depends on the difference in the specific heat of the gases, since $\Delta S = \Delta_1 S + \Delta_2 S$. This resolves the main part of Gibbs' paradox, which objected to having the same entropy change no matter the physicochemical properties of the two gases. In fact this is not so, if the effect of thermal radiation is appropriately accounted for. The established formulae also show that small variations in the difference of the specific heats of the two gases produce small variations in the predicted value of $\Delta S$, which complete the answer to the main part of the paradox.

What remains to be considered is the part of Gibbs' paradox that refers to the particular case in which the gases in the two chambers are the same and are at the same initial pressure and temperature. In this case the two gases will not expand as the partition is removed. This is because the number of gas particles that go from chamber 1 to chamber 2 after the partition removal will be equal to the number of gas particles that go from chamber 2 to chamber 1, the two gases being in thermal equilibrium with each other. Since the particles of the two gases are identical, the net effect of this particle exchange will be the same as if the partition was not removed at all. It is obvious then that in this case the partition removal cannot produce any entropy change at all, which answers the same-gas part of the paradox.

One may still wonder, however, why should the entropy change vanish in the case of two identical gases, whilst the slightest difference in their specific heat would produce the finite entropy jump $\Delta S = \Delta_1 S + \Delta_2 S$. There is no inconsistency here, either. The point is that the volumetric contribution $\Delta_1 S$ always applies in the case of two different gases –no mater how little their specific heats differ from each other, as long as they are not the same. The same contribution, however, is altogether absent in the case of two identical gases, since, as observed above, no expansion can take place in that case. It should be obvious then that one should not compare the entropy change due to a process that includes an expansion, with that of another process that does not.

We can also observe that if we confine our attention to the contribution $\Delta_2 S$, which is not due to the expansion, then the entropy change relevant to the case of two different gases will tend to zero as $c'_v$ and $c''_v$ tend to the same value. This is perfectly consistent with the fact that $\Delta S=0$ in the limit case in which the two gases are identical and answers the same-gas part of Gibbs' paradox concerning the alleged entropy jump inconsistency too.

It might still be objected that if we put a red gas in chamber 1 and a white gas in chamber 2 we would get a pink mixture once the partition is removed, even if the two gases are "*the same*" in that $c'_v = c''_v$. This could be seen as the evidence that the entropy of the system would increase, contrary to what we concluded above. Such a change in colour, however, is not a thermodynamic process. "Redness" and "whiteness" are not state variables of the system. Nor do they enter the state equations of the gases. As a consequence, no internal energy or entropy change can be produced in the system by the colour change of the two gases. No work or heat is absorbed in the process either. The gas colour change process that would certainly take place in the considered situation is a purely mechanical process; a consequence of the disordered distribution of the velocities of the gas particles. It also could be regarded as a demonstration of the thermal agitation of the gas particles, which is always active in any gas, even in thermal equilibrium, provided that $T \neq 0$.

But, is the pink state of the two gases more disordered than the initial red and white one? Perhaps. In some sense at least. However, as remarked in [6], disorder and macroscopic entropy are not always related.